\documentclass[twocolumn,twoside,slac_two]{revtex4}
\usepackage{graphicx}
\usepackage{fancyhdr}
\begin{document}

\title{The Connection between Radio and Gamma-ray Emission in Active Galactic Nuclei}

%

%

\author{Marcello Giroletti}
\affiliation{INAF Istituto di Radioastronomia, Bologna, Italy}

\author{Anita Reimer}
\affiliation{Leopold-Franzens-Universit\"at Innsbruck, Austria}

\author{Lars Fuhrmann}
\affiliation{Max-Planck-Institut f\"ur Radioastronomie, Bonn, Germany}

\author{Vasiliki Pavlidou}
\affiliation{California Institute of Technology, Owens Valley Radio
  Observatory}

\author{Joseph L.\ Richards}
\affiliation{California Institute of Technology, Owens Valley Radio
  Observatory}

\author{on behalf of the Fermi-LAT collaboration}
\noaffiliation

\begin{abstract}
Radio and gamma-ray emission from active galactic nuclei (AGN) are thought to
share a common origin, related to the ejection phenomena in the vicinity of
supermassive black holes. Thanks to its sensitivity, surveying capability, and
broad energy range, the Large Area Telescope (LAT) onboard the {\it Fermi
  Gamma-ray Space Telescope} has permitted us to discover and characterize a
huge number of extragalactic $\gamma$-ray sources. Similarly to what was found
by EGRET, these sources are typically associated with blazars, characterized by
significant radio emission and flat spectrum. The radio luminosity distribution
is extended over 7 orders of magnitudes, with flat spectrum radio quasars
clustered at higher powers and BL Lacs more scattered; the average spectral
index is consistent with $\alpha=0$, although a few remarkable sources have
$\alpha>0.5$ ($S(\nu) \propto \nu^{-\alpha}$).  A comparison of the radio flux
density and the gamma-ray photon flux is presented, although claims on its
significance require a detailed discussion and Monte Carlo simulations which
will be presented in a future paper.

\end{abstract}

\maketitle

\thispagestyle{fancy}

\section{Introduction}

Around 10\% of active galactic nuclei (AGN) are strong sources of radio
emission. This includes radio galaxies, radio quasars (flat or steep spectrum),
and BL Lac type objects. All these sources are generally referred to as radio
loud (RL) AGN, whereas flat spectrum radio quasars (FSRQ) and BL Lac objects
are collectively known as {\it blazars}. Interestingly, the vast majority of
identified extragalactic sources in the third EGRET catalog
\citep[3EG,][]{Hartman1999} belong to the blazar class.


RL AGN, and blazars in particular, are the most numerous and most luminous
class of extragalactic gamma-ray sources.  That these sources are bright in
both gamma-ray and radio suggests a connection between the emission processes
in the two energy bands.  Radio emission from blazars (and RL AGN in general)
is generally accepted to be synchrotron radiation emitted by relativistic
electrons, while the physical processes responsible for the $\gamma$-ray
emission are much less well constrained.  The presence of non-thermal
synchrotron emission implies the existence of a population of relativistic
electrons. In the presence of low energy seed photons and relativistic beaming,
Inverse Compton (IC) up-scattering of the photons is frequently invoked to
explain the $\gamma$-ray emission.


Indeed, in the well known {\it blazar sequence}
\citep{Fossati1998,Ghisellini1998,Donato2001}, it is proposed that the
synchrotron and inverse Compton mechanisms give rise to a connection between
the radio luminosity and the peak frequencies and relative intensities of the
characteristic two-humped spectral energy distribution (SED) of blazars.  On
the other hand, evidence for a direct correlation between radio and
$\gamma$-ray flux density or luminosity has been widely debated since the EGRET
era \citep[e.g.][]{Stecker1993,Salamon1994,Taylor2007,Bloom2008,Kovalev2009}
but not conclusively demonstrated, when all the relevant biases and selection
effects are considered \citep[e.g.][]{Muecke1997}.

The Large Area Telescope (LAT) onboard the {\it Fermi Gamma-ray Space
Telescope}, with its large field of view and unprecedented sensitivity, is now
putting us in the condition of a better understanding of the extragalactic
$\gamma$-ray source population. In anticipation of the launch of {\it Fermi},
large projects in the radio band have been undertaken
\citep[e.g.][]{Healey2007,Fuhrmann2007,Richards2009,Lister2009}. The results of
these projects can now be exploited to gain insights into the radio properties
of this population \citep[see also][]{Max-Moerbeck2009} and into the relation
between radio and gamma-ray properties.

In the present paper, we report in \S \ref{s.1} the main results obtained after
three months of sky-survey operation \citep[LAT Bright AGN Sample,
  LBAS,][]{Abdo2009a} and in \S\ref{s.2} some analysis of an expanded source
set based on a preliminary version of the 1 year catalog. Conclusions and
future plans are given in \S\ref{s.3}. We use a $\Lambda$CDM cosmology with $h
= 0.71$, $\Omega_\mathrm{m} = 0.27$ and $\Omega_\Lambda = 0.73$, where the
Hubble constant $H_0=100h$ km s$^{-1}$ Mpc$^{-1}$ and define radio spectral
indices $\alpha$ such that the flux density $S(\nu)\propto\nu^{-\alpha}$.

\section{Radio/gamma-ray connection in the LBAS\label{s.1}}

In the first three months of sky-survey operation, the LAT has revealed 125
non-pulsar bright sources at $|b|>10^\circ$ with test statistic greater than
100 ($TS>100$, corresponding to a detection significance of $\sim 10 \sigma$);
106 of these sources have high-confidence associations with known AGN, 10 have
lower-confidence associations, and only 9 remain unassociated. Therefore,
considering both high- and lower-confidence associations, the fraction of
high-latitude bright $\gamma$-ray sources associated with radio loud AGN is as
high as 93\%. While EGRET results suggested that RL AGN would dominate this
population, this degree of dominance is nonetheless amazing. More than 50\%
(96/181) of the high-latitude ($|b|>10^\circ$) sources in the 3EG catalog were
unassociated and it seems that the overwhelming majority of those were RL AGN!

\begin{figure}
\includegraphics[width=\columnwidth]{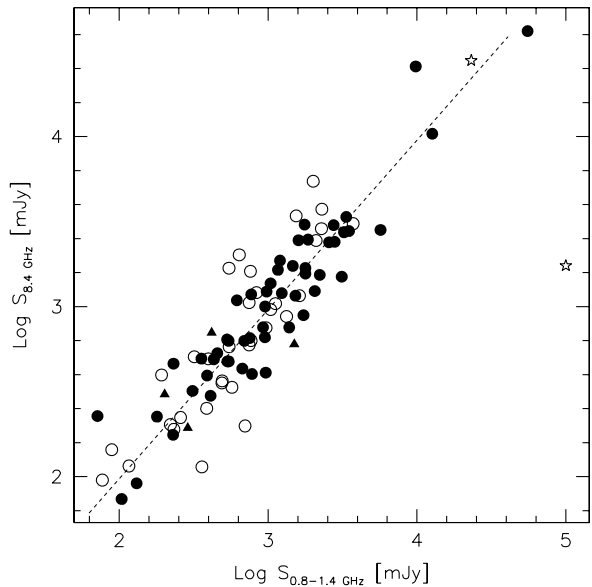}
\caption{\label{f.1} High vs.\ low frequency radio flux density for LBAS
  sources; the dashed line corresponds to $S_\mathrm{low}=S_\mathrm{high}$.  }
\end{figure}

Of the 106 LBAS sources, 104 are blazars, consisting of 58 FSRQs
\citep[including one narrow-line Seyfert~1,
J0948+0022,][]{Abdo2009b,Abdo2009d}; 42 BL Lacs; four blazars with unknown
classification; and two radio galaxies (Centaurus A and NGC 1275). The basic
radio properties can be obtained from the radio catalogs used for the
associations, i.e.\ CRATES \citep{Healey2007} and BZCat
\citep{Massaro2009}. CRATES provides 8.4 GHz VLA data for sources brighter than
$S_{4.8\,\mathrm{GHz}}=65\,$mJy, typically with subarcsecond resolution. BZCat
is a multi-frequency catalog giving low frequency radio data, typically from
the NVSS. Not surprisingly, given the flux and spectral selection criteria of
the samples, the LBAS sources are relatively bright ($98/106\sim92\%$ have
$S_{8.4}>100\,$mJy) and have flat spectral index ($\langle \alpha \rangle
=0.02\pm0.27$), which are typical signatures of compact self-absorbed
components.

\subsection{Low/high radio frequency comparison}

\begin{figure}
\center
\includegraphics[width=\columnwidth]{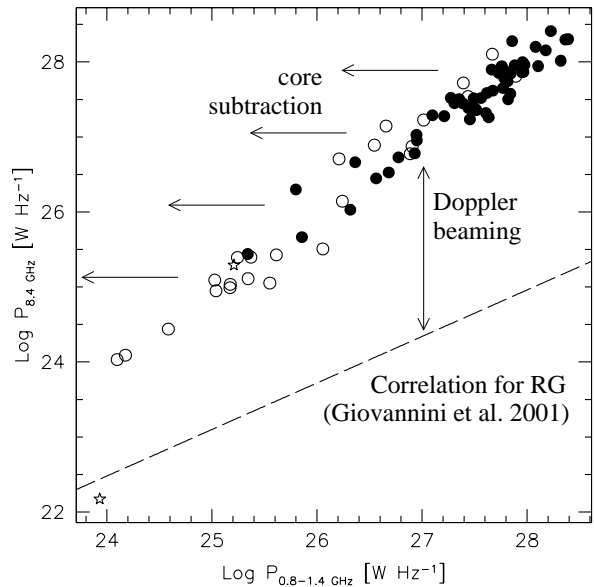}
\caption{\label{f.2} High vs.\ low frequency radio luminosity for LBAS sources;
  the dashed line shows the best fit for a sample of radio galaxies
  \citep{Giovannini2001}. Filled circles:\ FSRQs; open circles:\ BL Lacs;
  triangles:\ blazars of unknown type; stars:\ radio galaxies.  }
\end{figure}

In Fig.~\ref{f.1} and Fig.~\ref{f.2}, we show the comparison between the low
and high frequency radio flux density and luminosity for the LBAS sources. The
flux density plane (Fig.~\ref{f.1}) clearly reflects the flatness of the
spectral index, with all the points (but one) close to the
$S_\mathrm{low}=S_\mathrm{high}$ line; the outlying point represents Cen\,A,
the only source with a significant amount of extended radio emission dominating
the low frequency flux density. In the luminosity plot (Fig.~\ref{f.2}), the
common dependence on the square of the distance stretches the distribution even
more. Here, the interesting result is the relative distribution of the LBAS
sources with respect to a complete sample of radio galaxies \citep[dashed line,
  from][]{Giovannini2001}: except Cen\,A (which lies near the radio galaxy
region), all cores of LBAS sources are more luminous when compared to radio
galaxies of the same extended luminosity, which can be accounted for by a large
Doppler factor for the radio cores of $\gamma$-ray sources. Moreover, the low
frequency luminosity (generally taken as a measure of the extended, optically
thin emission) is contaminated by the nuclear emission for LBAS sources, so
that the actual extended emission luminosity would be lower than the observed
$P_\mathrm{low}$ (corresponding to a shift of the points in the direction of
the parallel arrows in Fig.~\ref{f.2}). The two main consequences are as
follows: (1) the Doppler beaming effect can actually be larger than apparent
from the figure and (2) $\gamma$-ray sources can be found even in blazars with
extended emission as low as $\sim10^{23}\,$W\,Hz$^{-1}$.

\subsection{Radio luminosity}

In terms of the radio luminosity $L_\mathrm{r}=\nu L(\nu)$ (calculated at
$\nu=8.4\,$GHz), the sources in the present sample with a measured redshift
span the range $10^{39.1}$ erg s$^{-1} < L_\mathrm{r} < 10^{45.3}$ erg
s$^{-1}$. 
BL Lacs and FSRQs are not uniformly distributed in this interval: the former
span a broad range of radio luminosities ($\log
L_{\mathrm{r,\,BL\,Lac}}\;[\mathrm{erg\,s}^{-1}] = 42.8\pm1.2$) while the
latter are more clustered at high radio luminosity ($\log
L_{\mathrm{r,\,FSRQ}}\;[\mathrm{erg\,s}^{-1}] = 44.4\pm0.6$).  Blazars of
unknown type have low-S/N optical spectra, so a redshift is generally not
available and their radio luminosities are not determined. Of the two radio
galaxies associated with objects in the LBAS, NGC~1275 is similar to the BL
Lacs ($L_\mathrm{r}=10^{42.2}$ erg s$^{-1}$), while Cen~A lies at the very low
end of the radio power distribution ($L_\mathrm{r}=10^{39.1}$ erg s$^{-1}$).

\begin{figure}
\center
\includegraphics[width=\columnwidth]{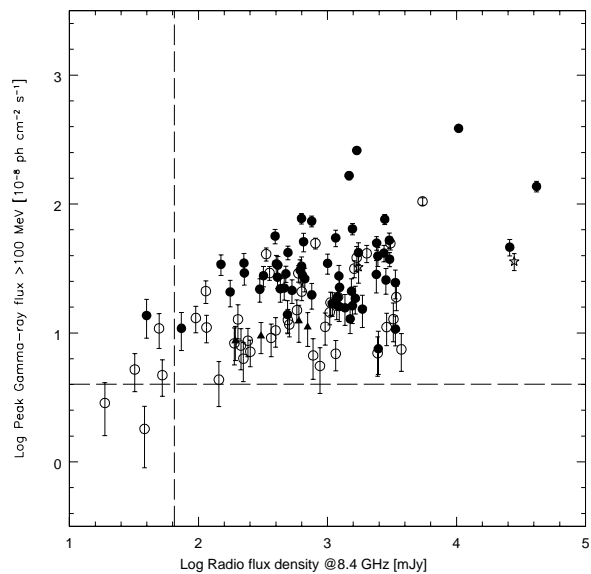}
\caption{\label{f.3} Peak gamma-ray flux vs.\ radio flux density at 8.4 GHz for
  LBAS sources; the dashed lines show the CRATES flux density limit and the
  typical LAT detection threshold in three months. Filled circles:\ FSRQs; open
  circles:\ BL Lacs; triangles:\ blazars of unknown type; stars:\ radio
  galaxies. }
\end{figure}

\subsection{Gamma-ray vs radio plane}

\begin{figure}
\center
\includegraphics[width=\columnwidth]{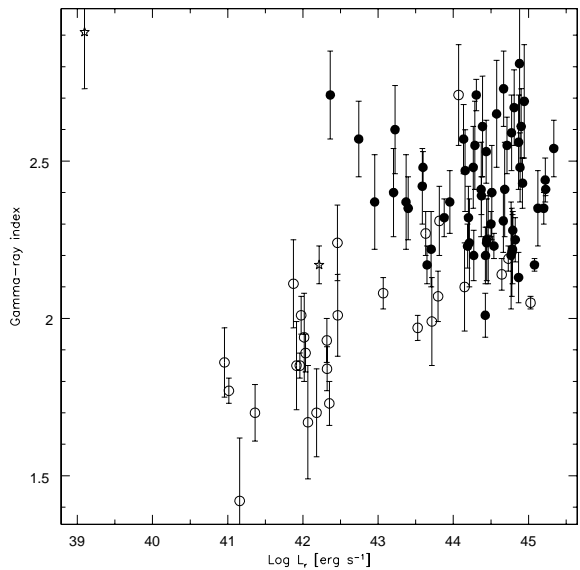}
\caption{\label{f.4} Gamma-ray photon index vs.\ radio luminosity for LBAS
  sources. Filled circles:\ FSRQs; open circles:\ BL Lacs; triangles:\ blazars
  of unknown type; stars:\ radio galaxies. The object in the top left corner is
  the gamma-ray source associated to Cen\,A.  }
\end{figure}

By combining data in the radio archives and the LAT measurements, it is
possible to compare the properties of the LBAS sources at low and high energy.
The relevant plots are shown in Fig.~\ref{f.3} and Fig.~\ref{f.4} and concern
the peak $\gamma$-ray flux vs.\ the radio flux density and the radio luminosity
vs.\ $\gamma$-ray spectral index planes, respectively. These plots are also
discussed in detail in \citet{Abdo2009a}.

The calculation of a simple Spearman's rank correlation coefficient $\rho$ for
the distribution in the flux-flux plane (Fig.~\ref{f.3}) of the 106 objects
yields $\rho=0.42$. However, the actual significance of this result can not be
assessed without Monte Carlo simulations, since biases and/or selection effects
modify the distribution of $\rho$ from the ideal case. Moreover, when FSRQs and
BL Lacs are considered separately, quite different results are obtained
($\rho_{\mathrm{FSRQ}}=0.19$, $\rho_{\mathrm{BL\,Lac}}=0.49$), so that it is
not possible to claim significant correlations at this stage. Furthermore, the
radio and $\gamma$-ray data are not simultaneous. The importance of this effect
needs to be taken into account, and the only method to address its impact is to
apply similar tests to a simultaneous sample, such as the one provided from the
F-GAMMA (OVRO 15 GHz, Effelsberg \& IRAM 30m 2-230 GHz) monitoring
programs. First results using these data indicate significant correlations
\citep[][Fuhrmann et al., Richards et al., in prep.]{Richards2009}.

The other relevant plot for the comparison of radio and $\gamma$-ray properties
is the radio luminosity vs.\ $\gamma$-ray spectral index plane (see
Fig.~\ref{f.4}).  The broad LAT energy range permits to readily reveal the
separation between BL Lacs and FSRQs, with FSRQs at largest $L_r$ and softer
indices and BL Lacs at lower $L_r$ and harder indices. As far as the two radio
galaxies are concerned, NGC\,1275 is similar to BL Lacs, while Cen~A is well
displaced, having a much softer $\gamma$-ray spectral index than other
low-power radio sources.

\section{Radio/gamma-ray connection with expanded source set}\label{s.2}

\begin{figure}
\center
\includegraphics[width=\columnwidth]{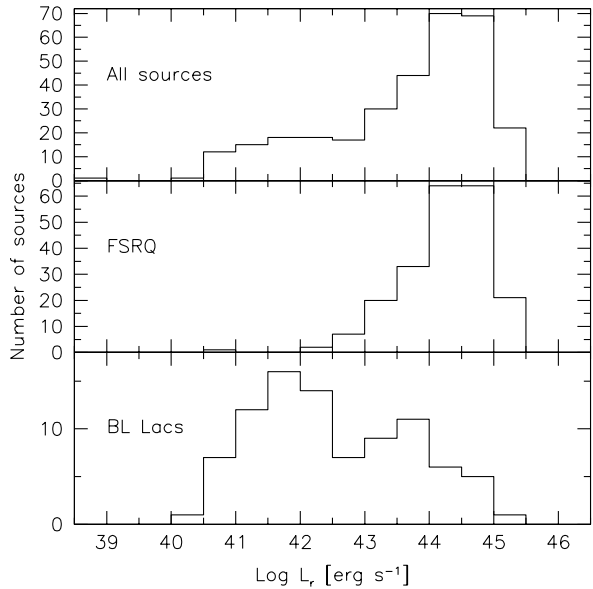}
\caption{\label{f.5} Preliminary histogram of radio luminosity distribution for
  AGN associated with LAT sources in the 1-yr catalog. }
\end{figure}

\begin{figure}
\includegraphics[width=\columnwidth]{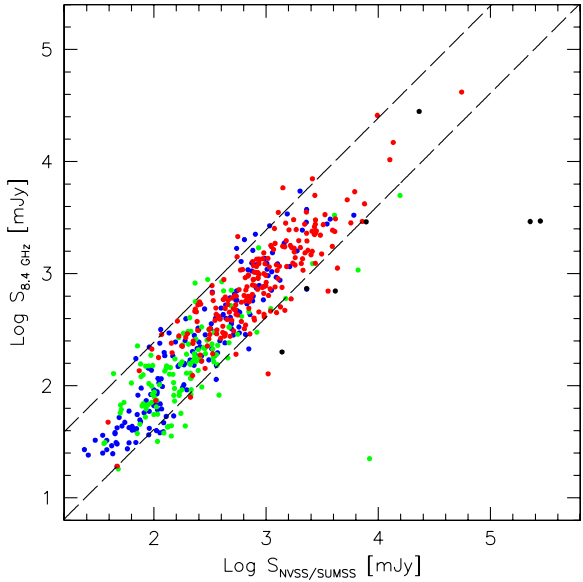}
\caption{\label{f.6} Preliminary distribution of AGN associated with 1-yr LAT
  sources in the high vs.\ low frequency radio flux density; the dashed lines
  correspond to $\alpha = \pm0.5$. Red circles: FSRQs; blue: BL Lacs; black:
  radio galaxies; green: other AGN. }
\end{figure}

\begin{figure}
\includegraphics[width=\columnwidth]{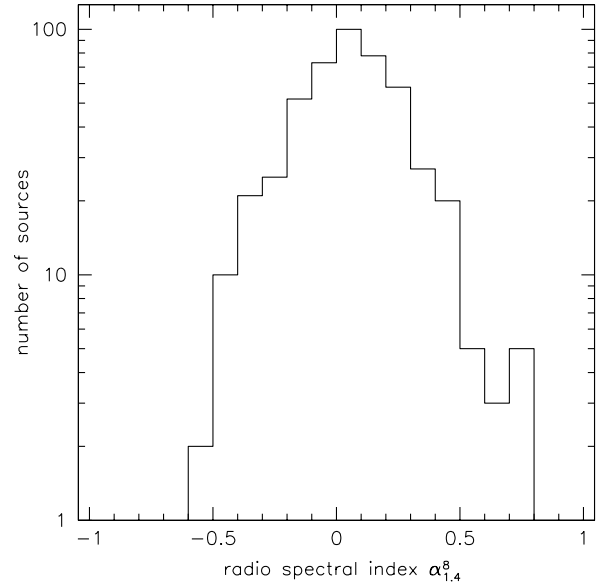}
\caption{\label{f.7} Preliminary histogram of radio spectral index distribution
  for AGN associated with LAT sources in the 1-yr catalog.  }
\end{figure}

With the continuation of the LAT operation in sky-survey mode, the number of
detected blazars is bound to increase. In the 1-year catalog under development
from the LAT team\footnote{The 1-year {\it Fermi} catalog is available online
  at http://fermi.gsfc.nasa.gov/ssc/data/access/lat/1yr\_catalog/}, more than
1000 sources have been detected and characterized \citep{Ballet2009}. In
particular, the {\it large field of view} of the LAT has allowed us to reach
weak gamma-ray sources all over the sky and the search for their possible radio
counterparts has required a huge amount of work \citep[see
  e.g.][]{Healey2009}. Indeed, while the LBAS results were based on sources
with $TS>100$ in three months of survey, the 1-year preliminary results
presented here are based on sources with $TS>25$ (significance about
$5\sigma$). As long as there is a significant overlap with the LBAS, the total
number of sources available for analysis becomes several times as large, and
the explored space of parameters is also larger. Thanks to the {\it great
  sensitivity} and {\it broad energy range} of the LAT, {\it Fermi} has already
been successful in revealing faint $\gamma$-ray sources and in the
characterization of their diverse photon indices.

It is therefore interesting to have a preliminary look at the topics discussed
in the previous section using the 1-year database.  The total radio luminosity
range of the associated sources spans seven orders of magnitude; the overall
distribution remains similar to that found for the LBAS sources. Also the
different distributions of FSRQs and BL Lacs are confirmed, with $\log
L_{\mathrm{r,\,FSRQ}}\;[\mathrm{erg\,s}^{-1}] = 44.2\pm0.7$ and $\log
L_{\mathrm{r,\,BL\,Lac}}\;[\mathrm{erg\,s}^{-1}] = 42.2\pm1.2$
(Fig.~\ref{f.5}). In particular, the BL Lacs remain spread over a wider
interval of radio luminosities, even with a hint of bimodality (note however
that the counts are smaller for the BL Lacs since their redshift is generally
more difficult to determine).

As far as the spectral properties are concerned, we show in Fig.~\ref{f.6} the
distribution of the associations in the high vs.\ low frequency radio flux
density and in Fig.~\ref{f.7} the corresponding histogram of the spectral index
distribution. It is readily apparent from both plots that the vast majority of
sources have a flat spectrum ($\langle \alpha \rangle = 0.06 \pm 0.23$),
including sources as weak as a few tens of mJy. However, a small but
significant tail of steeper spectrum sources is also present, which are
described in more detail in dedicated works, such as M87
\citep{Finke2009,Abdo2009c} and Cen\,A \citep{Cheung2009}, as well as in the
review by \citet{Cavazzuti2009}.

\begin{figure}
\includegraphics[width=\columnwidth]{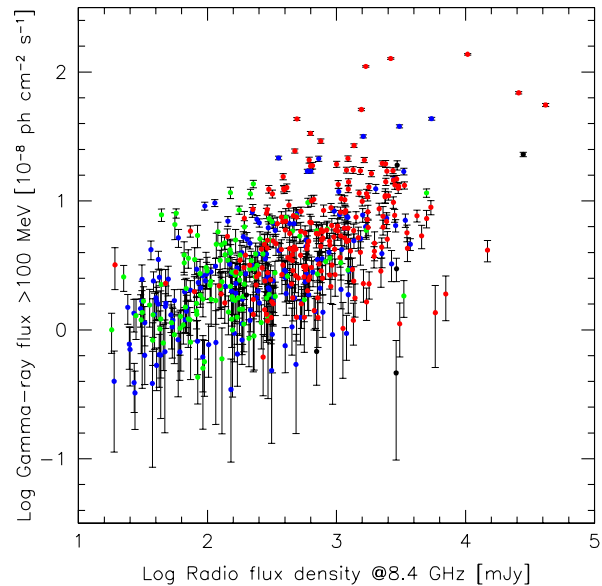}
\caption{\label{f.8} Preliminary distribution of 1-yr LAT sources in the
  gamma-ray flux vs.\ radio flux density plane. Red circles: FSRQs; blue: BL
  Lacs; black: radio galaxies; green: other AGN.}
\end{figure}

Finally, the radio flux density vs.\ mean $\gamma$-ray flux plane becomes more
populated, as shown in Fig.~\ref{f.8}. The Spearman's rank correlation
coefficient for the full sample changes to $\rho=0.57$. Of course, the remarks
noted above still apply: different source classes sample different regions of
the plane, data are not simultaneous, and observational biases could be
present. A more rigorous analysis is therefore necessary to claim the
significance and to discuss the physical implication of the observed
distribution. This will be the subject of a dedicated paper [Abdo et al., in
  preparation].

\section{Conclusions and outlook}\label{s.3}

Thanks to its unique capabilities, {\it Fermi} has revealed that the bright
gamma-ray extragalactic sky is dominated by radio loud AGN, and blazars in
particular, even more than it could be probed during the EGRET era. Therefore,
the $\gamma$-ray data from the LAT, as well as the radio data from large
ongoing projects, represent a great resource to constrain the physical
processes related to ejection phenomena in AGN. The preliminary analysis
presented in this work certainly needs a more detailed follow-up, although the
fraction and type of associated sources, the ranges of their core and extended
radio luminosity, and the distribution of the sources in the various
radio--$\gamma$-ray planes look already quite interesting.

In particular, the discussion of variability and the use of simultaneous data,
the comparison to radio data at higher frequencies \citep{Angelakis2009} and/or
resolution \citep{Ros2009}, and a deeper analysis with Monte Carlo simulations
are necessary before significant claims can be made. Finally, it is worthwhile
to remind that as we discover weaker $\gamma$-ray sources, we will need to
broaden our constraints on possible counterparts, including radio-faint BL Lacs
(HBLs), steep spectrum radio sources, and Seyfert galaxies.

\bigskip 
\begin{acknowledgments}

The authors would like to thank E.\ Angelakis, W.\ Max-Moerbeck, and
A. Readhead for useful discussions and comments on the presentation and
manuscript.  The $Fermi$ LAT Collaboration acknowledges support from a number
of agencies and institutes for both development and the operation of the LAT as
well as scientific data analysis. These include NASA and DOE in the United
States, CEA/Irfu and IN2P3/CNRS in France, ASI and INFN in Italy, MEXT, KEK,
and JAXA in Japan, and the K.~A.~Wallenberg Foundation, the Swedish Research
Council and the National Space Board in Sweden. Additional support from INAF in
Italy and CNES in France for science analysis during the operations phase is
also gratefully acknowledged. AR acknowledges support from a Marie Curie
International Reintegration Grant within the 7th European Community Framework
Programme. VP acknowledges suppor provided by NASA through Einstein
Postdoctoral Fellowship grant number PF8-90060 awarded by the Chandra X-ray
Center, which is operated by the Smithsonian Astrophysical Observatory for NASA
under contract NAS8-03060.
\end{acknowledgments}

\bigskip 

\begin{thebibliography}{99} 
\bibitem[Abdo et al.(2009a)]{Abdo2009a} Abdo, A.~A., Ackermann, M., Ajello, 
M., et al.\ 2009a, ApJ, 700, 597 

\bibitem[Abdo et al.(2009b)]{Abdo2009b} Abdo, A.~A., Ackermann, M., Ajello, 
M., et al.\ 2009b, ApJ, 699, 976 

\bibitem[Abdo et al.(2009c)]{Abdo2009c} Abdo, A.~A., Ackermann, M., Ajello, 
M., et al.\ 2009c, ApJ, 707, 55 

\bibitem[Abdo et al.(2009d)]{Abdo2009d} Abdo, A.~A., Ackermann, M., Ajello, 
M., et al.\ 2009d, ApJ, 707, 727 

\bibitem[Angelakis et al.(2009)]{Angelakis2009} Angelakis, E., Fuhrmann, 
L., Zensus, J.~A., et al.\ 2009, arXiv, arXiv:0910.0643 

\bibitem[Ballet et al.(2009)]{Ballet2009} Ballet, J., et al.\ 2009, (this
  conference)

\bibitem[Bloom(2008)]{Bloom2008} Bloom, S.~D.\ 2008, AJ, 136, 1533 

\bibitem[Cavazzuti et al.(2009)]{Cavazzuti2009} Cavazzuti, E., et 
al.\ 2009, (this conference)

\bibitem[Cheung et al.(2009)]{Cheung2009} Cheung, C.~C., et al.\ 2009, (this
  conference)

\bibitem[Donato et al.(2001)]{Donato2001} Donato, D., Ghisellini, G., 
Tagliaferri, G., \& Fossati, G.\ 2001, A\&A, 375, 739 

\bibitem[Finke et al.(2009)]{Finke2009} Finke, J., et al.\ 2009, (this
  conference)

\bibitem[Fossati et al.(1998)]{Fossati1998} Fossati, G., Maraschi, L., Celotti,
  A., et al.\ 1998, MNRAS, 299, 433

\bibitem[Fuhrmann et al.(2007)]{Fuhrmann2007} Fuhrmann, L., Zensus, J.~A.,
  Krichbaum, T.~P., et al.\ 2007, AIPC, 921, 249

\bibitem[Ghisellini et al.(1998)]{Ghisellini1998} Ghisellini, G., Celotti, A.,
  Fossati, G., et al.\ 1998, MNRAS, 301, 451

\bibitem[Giovannini et al.(2001)]{Giovannini2001} Giovannini, G., Cotton, 
W.~D., Feretti, L., Lara, L., \& Venturi, T.\ 2001, ApJ, 552, 508 

\bibitem[Hartman et al.(1999)]{Hartman1999} Hartman, R.~C., Bertsch, D.~L., 
Bloom, S.~D., et al.\ 1999, ApJS, 123, 79 

\bibitem[Healey et al.(2007)]{Healey2007} Healey, S.~E., Romani, R.~W., 
Taylor, G.~B., et al.\ 2007, ApJS, 171, 61 

\bibitem[Healey et al.(2009)]{Healey2009} Healey, S.~E., et al.\ 2009, (this
  conference)

\bibitem[Kovalev et al.(2009)]{Kovalev2009} Kovalev, Y.~Y., Aller, H.~D., 
Aller, M.~F., et al.\ 2009, ApJ, 696, L17 

\bibitem[Lister et al.(2009)]{Lister2009} Lister, M.~L., Aller, H.~D., 
Aller, M.~F., et al.\ 2009, AJ, 137, 3718 

\bibitem[Massaro et al.(2009)]{Massaro2009} Massaro, E., Giommi, P., Leto, 
C., et al.\ 2009, A\&A, 495, 691 

\bibitem[Max-Moerbeck et al.(2009)]{Max-Moerbeck2009} Max-Moerbeck, W., et
  al.\ 2009, arXiv:0912.3817 (these proceedings)

\bibitem[M\"ucke et al.(1997)]{Muecke1997} M\"ucke, A., Pohl, M., Reich, P., et
  al.\ 1997, A\&A, 320, 33

\bibitem[Richards et al.(2009)]{Richards2009} Richards, J.~L., et 
al.\ 2009, arXiv:0912.3780 (these proceedings)

\bibitem[Ros et al.(2009)]{Ros2009} Ros, E., for the MOJAVE collaboration 2009,
  arXiv, arXiv:0912.4644 (these proceedings)

\bibitem[Stecker et al.(1993)]{Stecker1993} Stecker, F.~W., Salamon, M.~H., \&
Malkan, M.~A.\ 1993, ApJ, 410, L71

\bibitem[Salamon \& Stecker(1994)]{Salamon1994} Salamon, M.~H., \& Stecker,
F.~W.\ 1994, ApJ, 430, L21

\bibitem[Taylor et al.(2007)]{Taylor2007} Taylor, G.~B., Healey, S.~E., 
Helmboldt, J.~F., et al.\ 2007, ApJ, 671, 1355 

\end{thebibliography}

\end{document}